\documentclass{article}
\usepackage{spconf,amsmath,graphicx}
\usepackage{mathtools}
\usepackage[ruled,vlined]{algorithm2e}
\usepackage{algorithmicx}

\title{3D Magnetic Inverse Routine for Single-Segment Magnetic Field Images}
\name{J. Senthilnath$^{1\star}$ \qquad Chen Hao$^{1}$ \qquad F. C. Wellstood$^{2,3\star}$
\thanks{This study is supported by the Machine Learning Guided Failure Analysis \& Diagnostic Capability Development for Next Generation 3D-IC Packaging at A*STAR via the IAF-PP by the Agency for Science, Technology and Research under Grant No. M23K8a0050. We acknowledge Ankur Singh for his efforts and insightful discussions at the outset of this work.}}
\address{$^{1}$Institute for Infocomm Research (I$^{2}$R), Agency for Science, Technology and Research (A*STAR), Singapore\\
$^{2}$Neocera Magma, Beltsville, MD, USA\\
$^{3}$Center for Nanophysics and Advanced Materials, Department of Physics,\\ University of Maryland, College Park, MD, USA\\
$^{\star}$Corresponding authors: J\_Senthilnath@i2r.a-star.edu.sg, well@umd.edu
}
\begin{document}
%\ninept
%
\maketitle
\begin{abstract}
In semiconductor packaging, accurately recovering 3D information is crucial for non-destructive testing (NDT) to localize circuit defects. This paper presents a novel approach called the \textbf{3D} \textbf{M}agnetic \textbf{I}nverse \textbf{R}outine (3D MIR), which leverages Magnetic Field Images (MFI) to retrieve the parameters for the 3D current flow of a single-segment. The 3D MIR integrates a deep learning (DL)-based Convolutional Neural Network (CNN), spatial-physics-based constraints, and optimization techniques. The method operates in three stages: i) The CNN model processes the MFI data to predict ($\ell/z_o$), where $\ell$ is the wire length and $z_o$ is the wire's vertical depth beneath the magnetic sensors and classify segment type ($c$). ii) By leveraging spatial-physics-based constraints, the routine provides initial estimates for the position ($x_o$, $y_o$, $z_o$), length ($\ell$), current ($I$), and current flow direction (positive or negative) of the current segment. iii) An optimizer then adjusts these five parameters ($x_o$, $y_o$, $z_o$, $\ell$, $I$) to minimize the difference between the reconstructed MFI and the actual MFI. The results demonstrate that the 3D MIR method accurately recovers 3D information with high precision, setting a new benchmark for magnetic image reconstruction in semiconductor packaging. This method highlights the potential of combining DL and physics-driven optimization in practical applications.
\end{abstract}
% Papers should be limited to 5 pages of technical content, including figures and references,
% and one optional 6th page containing only references/bibliography.

\section{Introduction}
\label{sec:intro}
The development of three-dimensional (3D) packaging in semiconductor manufacturing is crucial for advanced computer technology. However, non-destructive testing (NDT) of these structures becomes more challenging as package depth and the number of layers increase \cite{orozco2018non}. Most semiconductor manufacturers use vision-based analysis to examine advanced 3D packaging, such as 3D X-ray microscopy (3D XRM), for NDT. Although these popular 3D scanners provide high-resolution imaging, they can be time-consuming and require substantial storage space for large-scale scans \cite{gu2023breakthrough}. Therefore, there is a pressing need to either integrate deep learning (DL) into the reconstruction workflow to enhance scan speed or to use scanners that can significantly reduce scan time while exploiting imaging of physics. One such alternative is the use of magnetic field images (MFI), which can help decrease both scan time and storage requirements \cite{FFT1}. 

MFI allows the imaging of magnetic fields generated by currents through relatively thick insulating and metal layers. For MFI to be useful in testing 3D packages in semiconductor manufacturing, the MFI image needs to be converted to an image that represents the current flow. This conversion has commonly been done using a Fast Fourier Transform (FFT) \cite{FFT2} to achieve magnetic field inversion, which unfortunately does not allow full 3D resolution. Existing literature highlights work, on MFI image analysis for NDT \cite{vanderlinde2000localizing,hsiung2004failure}. For instance, the image difference analysis between the magnetic images from pins in functional components and those in failing components can identify defect locations. However, obtaining a reference image that corresponds to the exact scanned region in real-time is challenging. Most studies have concentrated on traditional image interpretation and have evaluated findings against X-ray \cite{vanderlinde2000localizing} and Acoustic Microscope in Confocal Scanning Acoustic Microscopy (C-SAM) \cite{hsiung2004failure} to detect anomalies. This approach often depends on experienced experts, which can significantly slow down the NDT process. To the authors' knowledge, there is currently no published literature that applies DL, spatial-physics-based constraints and optimization for 3D MFI representation.

In this paper, we introduce a novel approach called the 3D Magnetic Inverse Routine (3D MIR). This technique utilized CNN model, spatial-physics-based constraints, and optimization methods to convert magnetic images of current-carrying circuits into 3D representations of current flow. For a range of magnetic images, the 3D MIR method can achieve lateral and vertical resolution that are suitable for applications to microchip failure analysis. In summary, the key contributions of our work are three-fold:
\begin{enumerate}
    \item Develop two CNN models to analyze the MFI: 1 CNN model acts as a regression model to estimate the parameter $\beta=\ell/z$ and another CNN model acts as a classification model to classify the segment type $c$ ($x$-segment or $y$-segment).
    \item By leveraging spatial-physics-based constraints, our routine provides initial estimates for the current segment's position ($x_o$, $y_o$, $z_o$), length ($\ell$), current intensity ($I$), and the direction of current flow (positive or negative).
    \item An optimization process then adjusts the parameters ($x_o$, $y_o$, $z_o$, $\ell$, $I$) to minimize the difference between the reconstructed MFI and the actual MFI.
\end{enumerate}
We conducted extensive simulations to evaluate the proposed 3D MIR using single-segment magnetic images. The results demonstrate the method's effectiveness in reconstructing magnetic images, with the CNN providing initial estimates followed by parameter optimization.

\section{3D MIR Method}
\label{sec:method}
\begin{figure*}[htb]
\centering
\includegraphics[width=1\textwidth]{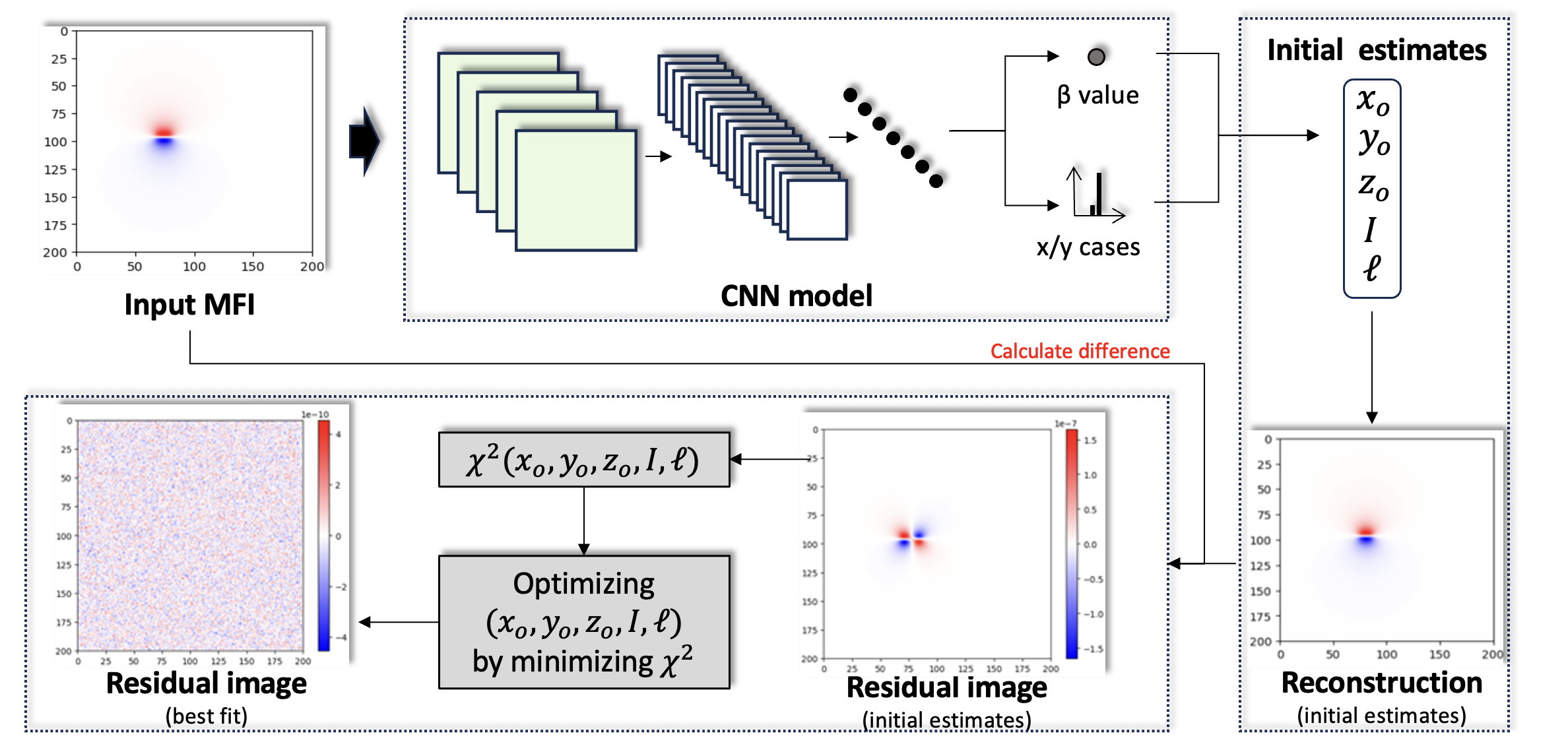}
\caption{Schematic overview of the 3D MIR. Notations - \((x_o, y_o, z_o)\) is the 3 dimensional coordinates of the start point of a current segment, \(\ell\) is the length of the segment, which carries a current intensity of \(I\) and \(\chi^2\) is the objective function defined for pixel-level reconstruction.}
\label{fig:3d mir}
\end{figure*}

Our 3D MIR method operates in three stages for reconstruction of magnetic images of current-carrying circuits into 3D representation of current flow. Figure \ref{fig:3d mir} shows the schematic representation of the 3D MIR method.

\subsection{Preliminaries}
\label{ssec:Pre}
Consider a current segment of length \(\ell\) that starts at \((x_o, y_o, z_o)\) and ends at \((x_o + \ell, y_o, z_o)\). It carries a current \(I\) in the \(+x\) direction. Assume that the width \(w\) and thickness \(d\) of the wire are negligibly small. For an arbitrary point \((x,y,z)\) in space, according to the Biot-Savart Law~\eqref{eq:biot_savart}, the magnetic field is given by~\eqref{eq:Bz_component}.
\begin{equation}
\mathbf{B} = \frac{\mu_o}{4\pi} \int \frac{\mathbf{J} \times \hat{\mathbf{r}}}{|\mathbf{r}|^2} \, dV
\label{eq:biot_savart}
\end{equation}
\noindent where $\mu_o$ is the magnetic permeability, $\mathbf{J}$ is the current density and the resultant magnetic flux density $\mathbf{B}$ is at position $\mathbf{r}$ in the 3D-space.

\noindent The z-component of $\mathbf{B}$ is
\begin{equation}
\begin{aligned}
&B_z(x, y, z) = \frac{\mu_od}{4\pi} \int_{-\infty}^\infty \int_{-\infty}^\infty  \\
&\frac{J_x(x', y') \cdot (y - y')}{[(x - x')^2 + (y - y')^2 + (z - z_o)^2]^{3/2}} \, dx' \, dy'
\end{aligned}
\label{eq:Bz_component}
\end{equation}
\noindent where $J_x$ is the $x$ component of the current density and we assume the current is carried in a layer of thickness $d$.

We define Peak-to-Peak (PP) as the distance between maximum 
$B_z$ and minimum $B_z$ and find,   
\begin{equation}
\begin{aligned}
&PP = y_{\text{max}} - y_{\text{min}} = \\
&\sqrt{\left(\frac{\ell}{2}\right)^2 + (z - z_o)^2}
\sqrt{-1 + \sqrt{1 + \frac{8(z - z_o)^2}{\left(\frac{\ell}{2}\right)^2 + (z - z_o)^2}}}
\end{aligned}
\label{eq:PP_formula}
\end{equation}
or:
\begin{equation}
\begin{aligned}
PP = |z - z_o| \sqrt{\left(\frac{\beta}{2}\right)^2 + 1}
\sqrt{-1 + \sqrt{1 + \frac{8}{\left(\frac{\beta}{2}\right)^2 + 1}}}
\end{aligned}
\label{eq:PP_reformula}
\end{equation}

We can rewrite the right side in terms of the unknown value of $z-z_o$ and the estimated value of \(\beta=\ell / |z-z_o|\)
~\eqref{eq:PP_reformula}, which gives
\begin{equation}
|z - z_o| =  \frac{PP}{ \sqrt{\left(\frac{\beta}{2}\right)^2 + 1}
\sqrt{-1 + \sqrt{1 + \frac{8}{\left(\frac{\beta}{2}\right)^2 + 1}}}}
\label{eq:l/z}
\end{equation}

To simplify the parameter form, we will use $z_o$ instead of $|z-z_o|$ in the following discussion, which represents the vertical depth beneath the magnetic sensor. 

\subsection{Parameters Initial Estimation}
\label{ssec:SC}
Given a MFI of a current segment, 3D MIR estimates the necessary parameters roughly from the MFI. Specifically, 3D-MIR estimates $\beta$ using a CNN model \cite{tan2019efficientnet}. It then identifies the maximum and minimum position of magnetic field intensity to get the PP distance and applies equation (\ref{eq:l/z}) to get current segment's depth ($z_o$). From $z_o$ and CNN predicted ($\beta$) we can get length ($\ell$) using,
\begin{equation}
\ell = z_o \beta
\label{l}
\end{equation}

Combining this information with the spatial-physics-based constraint of the MFI, we estimate the coordinates $x_o$ and $y_o$ of the current's starting point.

\begin{figure}[htb]
\centering
\begin{minipage}[b]{0.40\linewidth}
  \centering
  \centerline{\includegraphics[width=\linewidth]{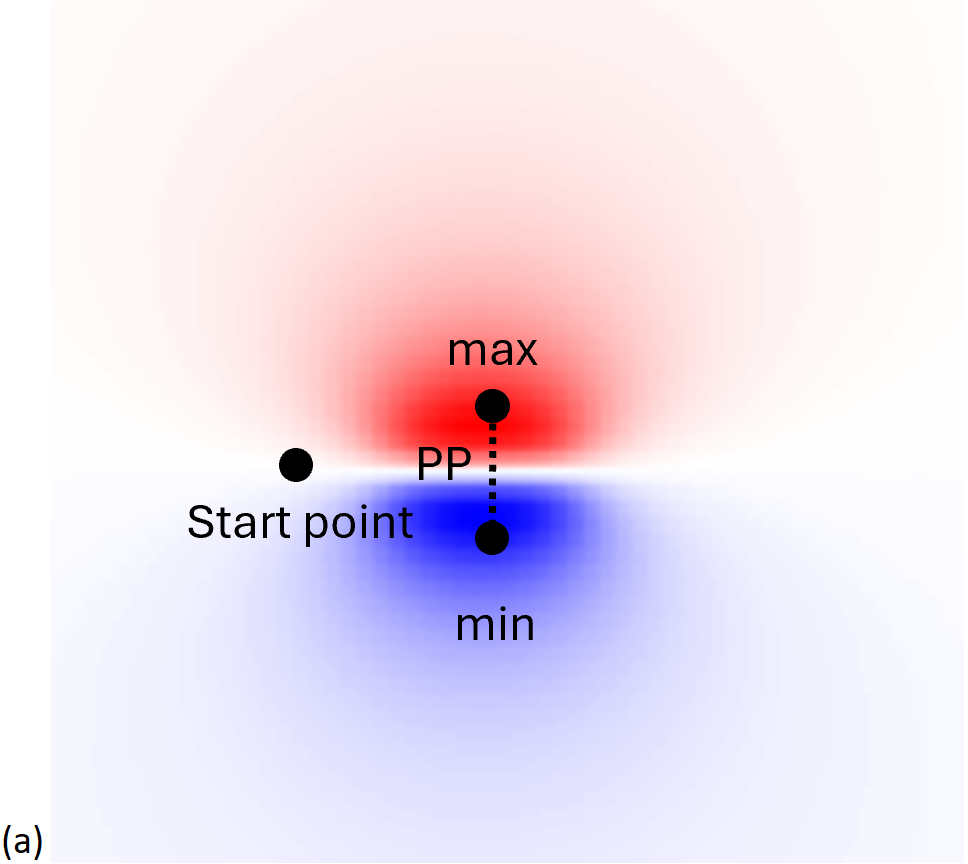}}
%  \vspace{1.5cm}
  %\centerline{}\medskip
\end{minipage}
%\hfill
\begin{minipage}[b]{0.40\linewidth}
  \centering
  \centerline{\includegraphics[width=\linewidth]{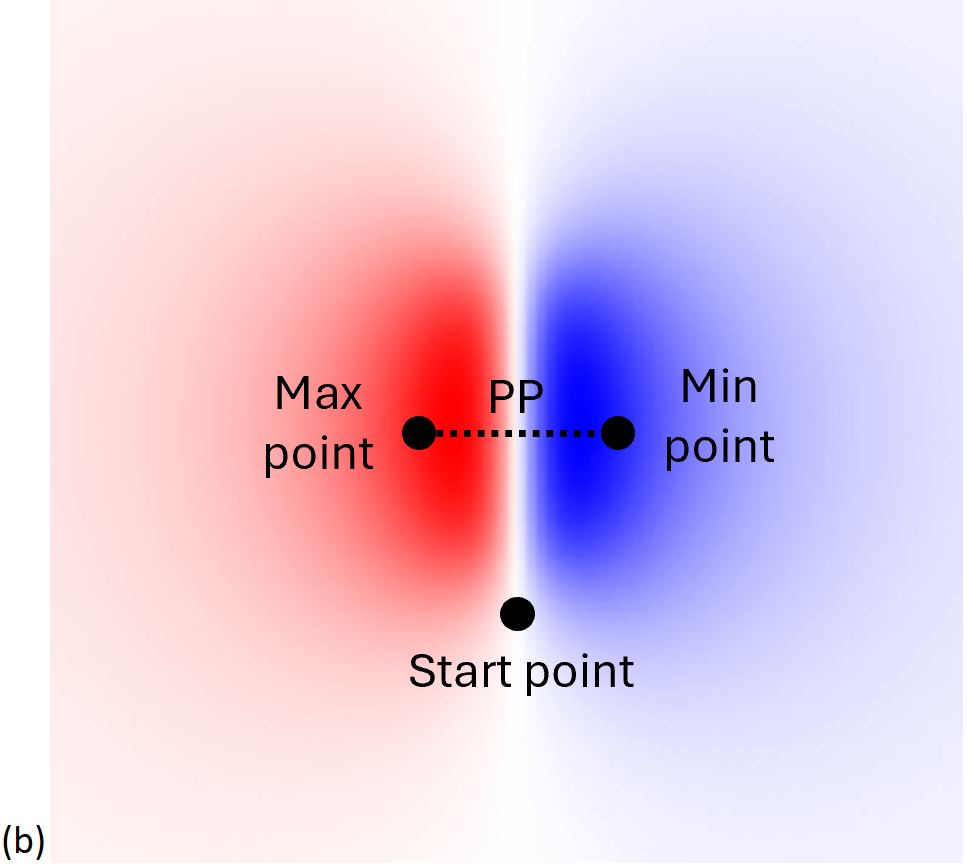}}
%  \vspace{1.5cm}
%  \centerline{}\medskip
\end{minipage}
\caption{PP distance for single-segment MFI data: (a) positive $x$-current, (b) positive $y$-current}
\label{fig:Twosamples}
\end{figure}

Next, a second CNN model is applied to the MFI image to classify the segment type ($x$ or $y$). If it is an $x$ segment (see Figure ~\ref{fig:Twosamples}(a)), we use the equation
\begin{equation}
\begin{aligned}
&y_o=\frac{y_{\text{max}} + y_{\text{min}}}{2},
&x_o=x_{\text{max}} - \frac{\ell}{2}
\end{aligned}
\label{x_situation}
\end{equation}

\noindent For a $y$ segment, we use (see Figure ~\ref{fig:Twosamples}(b)).
\begin{equation}
\begin{aligned}
&x_o= \frac{x_{\text{max}} + x_{\text{min}}}{2},
&y_o = y_{\text{max}} - \frac{\ell}{2}
\end{aligned}
\label{y_situation}
\end{equation}

\noindent Finally, the current intensity $I$ can be estimated using
%We use a machine learning model to recognize the form of MFI. 
\begin{equation}
\begin{aligned}
|I| \approx \frac{B_z^{\text{max}}}{\frac{\mu_o}{4\pi} \left( \frac{\ell}{2z} \right) \left( \frac{1}{\sqrt{\left(\frac{\ell}{2}\right)^2 + 2z^2}} \right)}
\end{aligned}
\label{I}
\end{equation}

\subsection{Parameter Optimization}
\label{ssec:Opt}
In section 2.2 we described the workflow for initial estimates of the parameters. In this section 3D MIR will optimize the parameters to achieve better magnetic image reconstruction.

The z-component of the magnetic intensity at the position \((x, y)\) can be calculated using (for an $x$-segment),
\begin{equation}
\begin{aligned}
B^{\text{test}}_z(x, y) &= \frac{\mu_o I}{4 \pi} \left( \frac{y - y_o}{(y - y_o)^2 + z^2} \right) \\
&\quad \times \Bigg( 
\frac{x_o + \ell - x}{\sqrt{(x - x_o - \ell)^2 + (y - y_o)^2 + z^2}} \\
&\quad - \frac{x_o - x}{\sqrt{(x - x_o)^2 + (y - y_o)^2 + z^2}} 
\Bigg)
\end{aligned}
\label{recon:x}
\end{equation}

\noindent For a $y$-segment
\begin{equation}
\begin{aligned}
B^{\text{test}}_z(x, y) &= -\frac{\mu_o I}{4 \pi} \left( \frac{x - x_o}{(x - x_o)^2 + z^2} \right) \\
&\quad \times \Bigg( 
\frac{y_o + \ell - y}{\sqrt{(y - y_o - \ell)^2 + (x - x_o)^2 + z^2}} \\
&\quad - \frac{y_o - y}{\sqrt{(x - x_o)^2 + (y - y_o)^2 + z^2}} 
\Bigg)
\end{aligned}
\label{recon:y}
\end{equation}

The objective cost function \(\chi^2\) is defined for pixel-level reconstruction
\begin{equation}
\begin{aligned}
\chi^2(x_o, y_o, z, \ell, I) &= \sum_{i=1}^{N^{\text{image}}} \sum_{j=1}^{N^{\text{image}}} \Bigg[ \frac{\left(B_{ij}^{\text{data}} - B_z^{\text{test}}(x_j, y_j)\right)^2}{\sigma_B^2} \Bigg]
\end{aligned}
\label{chi_square}
\end{equation}

To optimize the objective function \(\chi^2\) with respect to $x_o$, $y_o$, $z_o$, $\ell$ and $I$ the 3D MIR employs Nelder-Mead method \cite{nelder} to minimize iteratively, thereby reducing the difference between the reconstructed MFI $(B_z^{test})$ and the actual MFI $(B_z^{data})$. The converged minimum \(\chi^2\) value, along with the optimal fit parameters \((x_o, y_o, z_o, \ell, I)\) and best fit image $B_z^{\text{test}}$ are recorded. The residual image (best fit) shows hidden noise by reconstructing an image that closely matches the actual MFI (Figure \ref{fig:3d mir}). The 3D MIR for single-segment MFI is summarized in Algorithm \ref{alg:3D MIR} 

%%%%%%%%%%%%%%%% Algorithm
\begin{algorithm}[t]
%\SetAlgoLined
%\begin{algorithmic}[]

\noindent \textbf{Input:} A single current segment MFI $\mathcal{P}$, CNN model $M_1$ for $\beta$ prediction, CNN model $M_2$ for $x/y$ class prediction.

\textbf{Output:} Parameters describing the current segment in the MFI $(x_o, y_o, z_o, \ell, I)$.\\

\BlankLine
\tcc{Parameters Initial Estimation} 

1) 
\(\beta = M_1(\mathcal{P})\)\ , \(c = M_2(\mathcal{P})\) \;

2) From $\mathcal{P}$ get the maximum value of magnetic field \(B_z^{max}=max(\mathcal{P})\), the maximum and minimum coordinates $(x_{min},y_{min})$ $(x_{max}, y_{max})$, \;

3)
\If{$c$ is a $x$-segment}{
    $PP = \left|y_{\text{min}} - y_{\text{max}}\right|$ \;
    calculate $z_o$ by \eqref{eq:l/z}\;
    calculate $\ell$ by \eqref{l}\;
    calculate $x_o, y_o$ by \eqref{x_situation}\;
}
\ElseIf{$c$ is a $y$-segment}{
    $PP = \left|x_{\text{min}} - x_{\text{max}}\right|$ \; 
    calculate $z_o$ by \eqref{eq:l/z}\;
    calculate $\ell$ by \eqref{l}\;
    calculate $x_o, y_o$ by \eqref{y_situation}\;
}

4) Calculate $I$ by \eqref{I}\;
\BlankLine
\tcc{MFI reconstruction by initial estimates of the parameters}

5)
\If{$c$ is a $x$-segment}{
    Reconstruct the magnetic field at $(x, y)$ by \eqref{recon:x}
}
\ElseIf{$c$ is a $y$-segment}{
    Reconstruct the magnetic field at $(x, y)$ by \eqref{recon:y}
}

\BlankLine
\tcc{Parameter Optimization}

6) Minimize objective function ($\chi^2$) \eqref{chi_square} by Nelder–Mead method to optimize the parameters.

% \For {\textbf{x}$_s$ in \textbf{X}}{
% \textbf{v}$_{i}^{s}$\texttt{<-}\textbf{x}$_s$\;

% \For {$l = 1$ to $n$}{
%    Compute CD using (\ref{eq:eq9})-(\ref{eq:eq14}) to obtain $W^l$\;
   
%  Perform feedforward pass using $W^l$\; 
  
%   Extract features 
%   $\tilde{\textbf{x}}_{s}^{l}$ from layer \textbf{h}$^l$\;
% }
% Using extracted feature  $\tilde{\textbf{x}}_{s}^{l}$ predict $n_c$ (\ref{eq:eq16})\;

% Perform clustering with $n_c$ on $\tilde{\textbf{x}}_{s}^{l}$ using (\ref{eq:eq17})-(\ref{eq:eq20})\;

%  Evaluate $\eta$ using (\ref{eq:eq21})\;
% }

\textbf{return:}  $x_o, y_o, z_o, \ell, I$
%\end{algorithmic}
\caption{3D MIR for single-segment MFI}
\label{alg:3D MIR}
\end{algorithm}

\section{RESULTS}
\label{sec:results}
A comprehensive analysis has been conducted to evaluate the performance of the 3D MIR on single-segment MFI.

\subsection{Dataset and Implementation}
\label{ssec:datadesc}
\noindent \textbf{Data description.} A total of 6100 simulated magnetic images were segregated into three datasets for training, validation and testing. For training the CNN model, 5000 simulated magnetic images were used and 600 images were used for validation during each epoch of the training phase. Each image is a simulated MFI of either an $x$ or $y$ current segment, with added noise to mimic real MFIs. The remaining 500 magnetic images (test dataset) were reserved for evaluating the model’s performance in predicting $\beta=\ell/z$ and for classifying the segment type ($c$). 
% \begin{figure}[htb]
% \begin{minipage}[b]{0.48\linewidth}
%   \centering
%   \centerline{\includegraphics[width=\linewidth]{figures/beta loss.png}}
% %  \vspace{1.5cm}
% %  \centerline{(a) Regression model loss}\medskip
% \end{minipage}
% \hfill
% \begin{minipage}[b]{0.48\linewidth}
%   \centering
%   \centerline{\includegraphics[width=\linewidth]{figures/class model loss.png}}
% %  \vspace{1.5cm}
% %  \centerline{(b) Classification model loss}\medskip
% \end{minipage}

% \centering % 
% \includegraphics[width=0.48\linewidth]{figures/beta prediction performance.png}
% %  \vspace{1.5cm}
% %\centerline{(c) Actual vs predicted $\beta$ value}\medskip
% \caption{The training loss and $\beta$ prediction using CNN model: (a) Regression model loss, (b) Classification model loss, (c) Actual vs predicted $\beta$ value.}
% \label{fig:Two model logging}
% %
% \end{figure}

\begin{figure}[htb]
\begin{minipage}[b]{0.327\linewidth}
  \centering
  \centerline{\includegraphics[width=\linewidth]{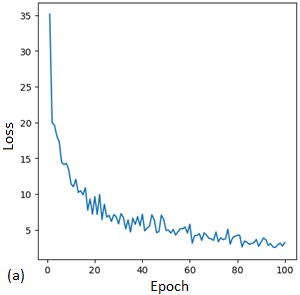}}
%  \vspace{1.5cm}
%  \centerline{(a) Regression model loss}\medskip
\end{minipage}
\hfill
\begin{minipage}[b]{0.327\linewidth}
  \centering
  \centerline{\includegraphics[width=\linewidth]{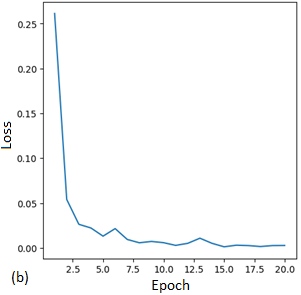}}
%  \vspace{1.5cm}
%  \centerline{(b) Classification model loss}\medskip
\end{minipage}
\hfill
\begin{minipage}[b]{0.327\linewidth}
  \centering
\centerline{\includegraphics[width=\linewidth]{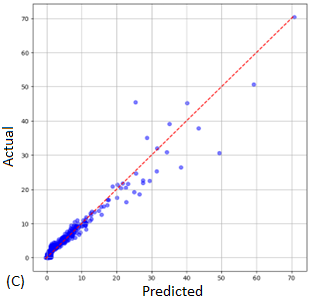}}
%  \vspace{1.5cm}
%\centerline{(c) Actual vs predicted $\beta$ value}\medskip
\end{minipage}
\caption{The training loss and $\beta$ prediction using CNN model: (a) Regression model loss, (b) Classification model loss, (c) Actual vs predicted $\beta$ value.}
\label{fig:Two model logging}
\end{figure}

\begin{figure}[ht!]
  \centering
  %\centerline{\includegraphics[width=\linewidth]
  \includegraphics[width=\linewidth]
  {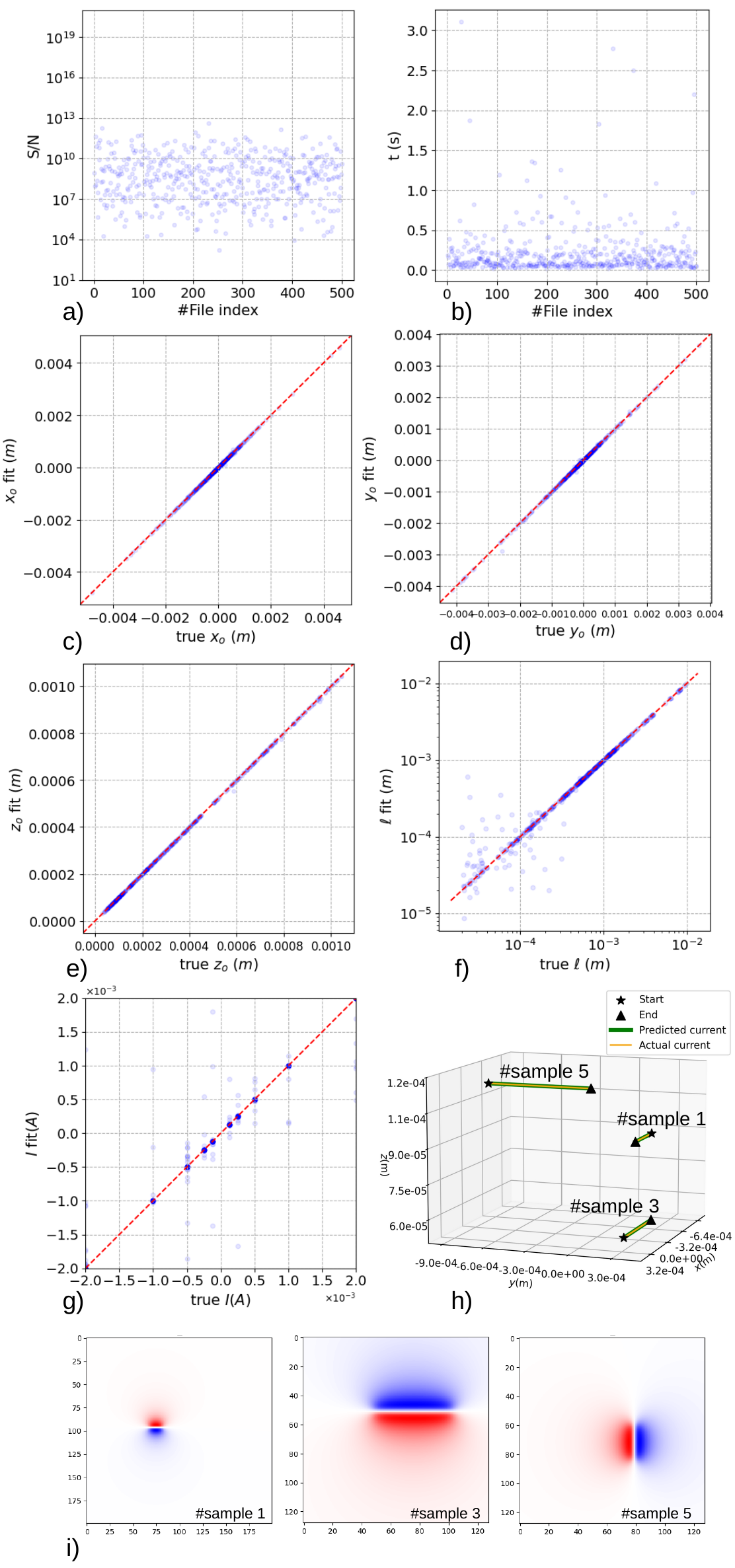}
\caption{Performance analysis of 500 test magnetic images: a) Signal-to-noise (S/N) distribution, b) Time for parameter optimization, c) Actual vs predicted $x_o$, d) Actual vs predicted $y_o$, e) Actual vs predicted $z_o$, f) Actual vs predicted $\ell$, g) Actual vs predicted $I$, h) Selected three samples showing predicted vs actual current segment in 3D space, i) MFI of the three samples reconstructed images.}
\label{fig:result1}
\end{figure}
\noindent \textbf{Implementation details.} A single-segment magnetic image has five parameters ($x_o$, $y_o$, $z_o$, $\ell$, $I$). The proposed method extracts initial estimates of these five parameters. We implemented our 3D MIR code in PyTorch and ran our experiments on NVIDIA RTX A6000 GPUs. To begin, the $\beta$ and $c$ values are predicted using the CNN model. We apply the EfficientNetB0 backbone \cite{tan2019efficientnet}, which is implemented in PyTorch \cite{Ansel_PyTorch_2_Faster_2024} for two independent processing steps: predicting $\beta$ using a regression approach (where the last layer features a linear layer with a single output) and classifying $c$ using classification approach (where the last layer features a linear layer with two output features followed by a softmax function). Furthermore, to optimize the five parameters, we applied the Nelder-Mead optimization method implemented which is available in the SciPy Python package \cite{2020SciPy-NMeth}. 

% \begin{figure}[htb]
% \centering
% \centerline{\includegraphics[\linewidth]{figures/training.pdf}}
% % \begin{minipage}[b]{0.49\linewidth}
% %   \centering
% %   \centerline{\includegraphics[width=\linewidth]{figures/beta loss.png}}
% % %  \vspace{1.5cm}
% % %  \centerline{(a) Regression model loss}\medskip
% % \end{minipage}
% % \hfill
% % \begin{minipage}[b]{0.49\linewidth}
% %   \centering
% %   \centerline{\includegraphics[width=\linewidth]{figures/class model loss.png}}
% % %  \vspace{1.5cm}
% % %  \centerline{(b) Classification model loss}\medskip
% % \end{minipage}

% % \centering % 
% % \includegraphics[width=0.5\linewidth]{figures/beta prediction performance.png}
% % %  \vspace{1.5cm}
% % %\centerline{(c) Actual vs predicted $\beta$ value}\medskip
% % \caption{The training loss and $\beta$ prediction using CNN model: (a) Regression model loss, (b) Classification model loss, (c) Actual vs predicted $\beta$ value}
% % \label{fig:Two model logging}
% %
% \end{figure}

% \begin{figure}[htb]
% \centering
% \includegraphics[width=0.48\textwidth]{figures/training.pdf}
% \caption{Schematic overview of the 3D MIR}
% \label{fig:Two model logging}
% \end{figure}

\subsection{CNN Model Performance}
\label{ssec:classify}
Figures~\ref{fig:Two model logging}(a) and (b) illustrate the convergence of training loss for the regression model and classification model respectively, while Figure~\ref{fig:Two model logging}(c) presents the prediction performance of the regression model. The classification model achieved an accuracy of 100\% in distinguishing between the $x$ and $y$ segments of the test MFI data.

The predicted value of $\beta$ and segment type ($c$) are used further to estimate the parameters discussed in Section 2.2, leading to initial estimates of all 5 parameters for each magnetic image. These initial parameter estimates are then refined using a Nelder–Mead optimizer to determine the best-fit values for the magnetic image reconstruction.

\subsection{Parameter Evaluation}
\label{ssec:LatVer}
\begin{figure}[ht]
  \centering
  \centerline{\includegraphics[width=\linewidth]{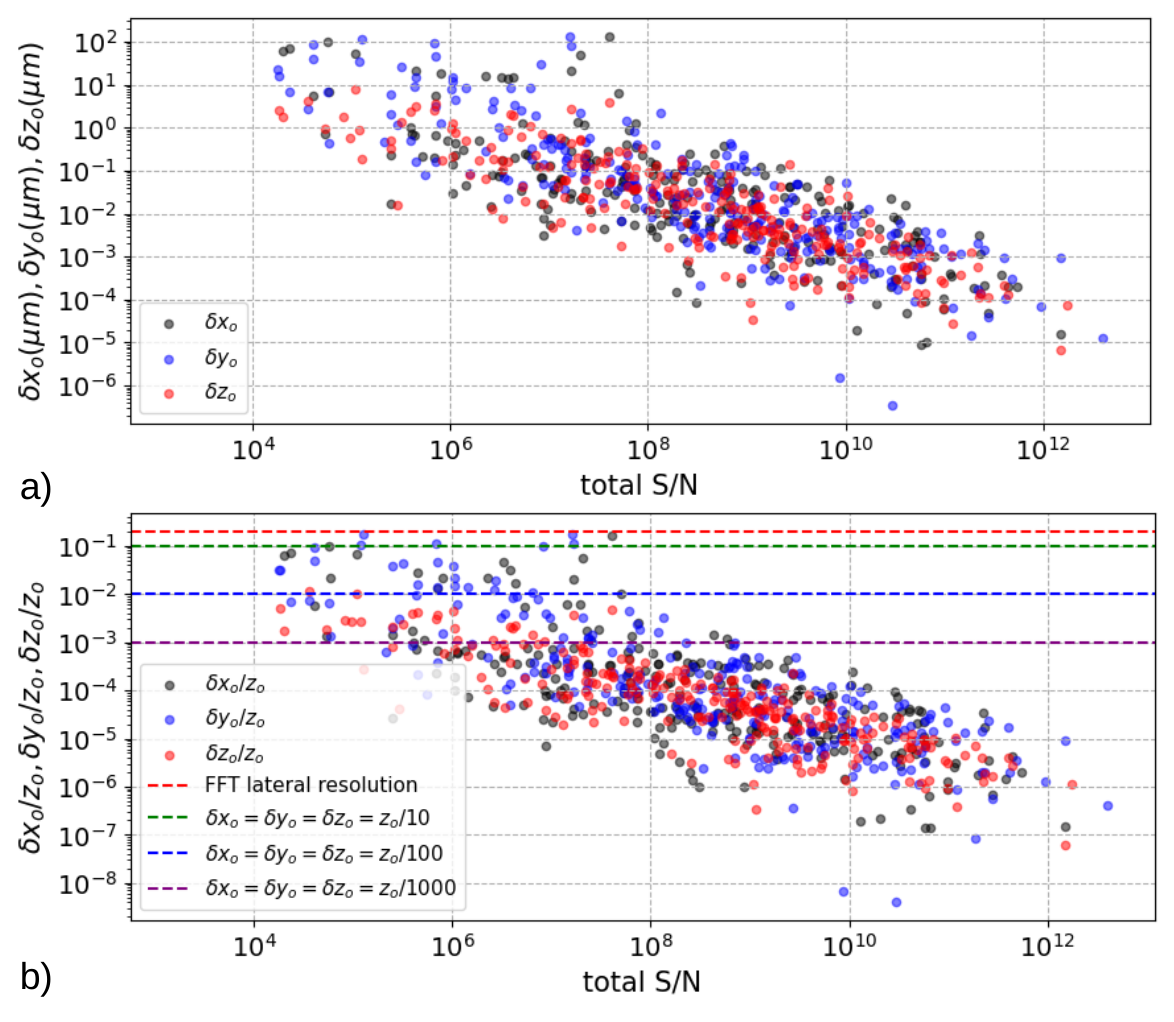}}
\caption{Error margins of 500 testing samples of magnetic images: a) absolute margins in $\mu$m, b) normalized margins by $z_o$ distance.}
\label{fig:result2}
\end{figure}

% In Figure 4, we evaluate how closely the estimated parameters of current align with actual current parameters through linear fitting. In Figure 5, we analyze the parameter differences and their resolution across different segments, which further validates the robustness of our method.
We applied the 3D MIR method to the 500 test MFI dataset and obtained 500 sets of optimized parameters. Each point in Figures ~\ref{fig:result1} and \ref{fig:result2} corresponds to one MFI image. Figure ~\ref{fig:result1}(a) shows the signal-to-noise (S/N) distribution across the 500 MFI data images, indicating that some images have high S/N ratios. Figure ~\ref{fig:result1}(b) presents the computation time (in seconds) required for the parameter optimization of each magnetic image. Figures ~\ref{fig:result1}(c), (d), (e), (f) and (g) show the scatter plots comparing actual MFI with predicted MFI data ($x_o, y_o, z_o, \ell$, $I$), where we can observe strong alignment along the diagonal (representing good agreement), except in cases when $\ell$ is too short. Figure ~\ref{fig:result1}(h) shows actual MFI with predicted MFI current segment in 3D space for three of the selected test magnetic images, reconstructed MFI images for these selected magnetic images are shown in Figure~\ref{fig:result1}(i).

Figure~\ref{fig:result2}(a) shows the error for $\delta x_o$, $\delta y_o$, and $\delta z_o$ are the difference between predicted and actual values in micrometers ($\mu$m), which decrease as S/N increases, indicating higher accuracy at higher S/N levels. Figure~\ref{fig:result2}(b) shows normalized errors ($\delta x_o/z_o$, $\delta y_o/z_o$, $\delta z_o/z_o$), which also improves with increasing S/N. The red dashed line indicates the best lateral spatial resolution achieved by Fast Fourier Transform (FFT) at high S/N, highlighting that 3D MIR can outperform FFT \cite{chatraphorn2001noise, chatraphorn2002relationship}.

\section{Conclusion}
\label{sec:refs}
In this paper, we propose the 3D Magnetic Image Routine (3D MIR) to estimate parameters in single-segment magnetic images. The 3D MIR method combines CNN, spatial-physics-based constraint and optimization techniques for parameter optimization. The trained CNN model is applied to 500 test magnetic images with varying wire lengths, from short to long. The parameter optimization yielded successful results through both quantitative and qualitative analysis in all cases. The integration of DL for extracting 3D information from magnetic images presents a novel approach to semiconductor packaging. Although this study focuses on a single-segment, there is potential for expanding the method to include multiple segments in future directions. Furthermore, the approach will be tested on real MFI data. 

% References should be produced using the bibtex program from suitable
% BiBTeX files (here: strings, refs, manuals). The IEEEbib.bst bibliography
% style file from IEEE produces unsorted bibliography list.
% -------------------------------------------------------------------------
%\vfill\pagebreak 

\newpage
\bibliographystyle{IEEEbib}
\bibliography{3DMIRrefs}

\end{document}